\documentclass[10pt,journal,compsoc]{IEEEtran}

\ifCLASSOPTIONcompsoc
  \usepackage[nocompress]{cite}
\else
  \usepackage{cite}
\fi

\usepackage{multirow}
\usepackage{subcaption}
\usepackage{url}
\usepackage{makecell}
\usepackage{amsmath}
\usepackage{amssymb}
\usepackage{mathtools}
\usepackage{booktabs}
\usepackage{algorithm}
\usepackage{algorithmic}
\newtheorem{theorem}{Theorem}[section]

\newtheorem{definition}[theorem]{Definition}
\newtheorem{corollary}[theorem]{Corollary}
\newtheorem{lemma}[theorem]{Lemma}

\ifCLASSINFOpdf
\usepackage[pdftex]{graphicx}
\graphicspath{{../pdf/}{../jpeg/}}
\DeclareGraphicsExtensions{.pdf,.jpeg,.png}
\else
\fi

\begin{document}
%
\title{Boosting Factorization Machines via Saliency-Guided Mixup}

\author{Chenwang~Wu,
        Defu~Lian,
        Yong~Ge,
        Min~Zhou,
        ~Enhong~Chen,~\IEEEmembership{Senior~Member,~IEEE,}
        and Dacheng Tao,~\IEEEmembership{Fellow,~IEEE,}
\IEEEcompsocitemizethanks{\IEEEcompsocthanksitem C. Wu is with the School of Data Science, University of Science and Technology of China, Hefei, Anhui 230000, China.\protect\\
E-mail: wcw1996@mail.ustc.edu.cn.
\IEEEcompsocthanksitem D. Lian and E. Chen are with the Anhui Province Key Laboratory of Big Data Analysis and Application, School of Computer Science and Technology, University of Science and Technology of China, Hefei, Anhui
230000, China.\protect\\
E-mail: \{liandefu, cheneh\}@ustc.edu.cn.
\IEEEcompsocthanksitem Yong Ge is with the Eller College of Management, University of Arizona.\protect\\
E-mail: yongge@arizona.edu.
\IEEEcompsocthanksitem Min Zhou is with the Huawei Noah's Ark Lab.\protect\\
E-mail: zhoumin27@huawei.com.
\IEEEcompsocthanksitem Dacheng Tao is with the JD Explore Academy in JD.com, China and the University of Sydney, Australia.\protect\\
E-mail: dacheng.tao@gmail.com.
\IEEEcompsocthanksitem Corresponding author: Defu Lian.}
\thanks{Manuscript received April 19, 2005; revised August 26, 2015.}}

\markboth{Journal of \LaTeX\ Class Files,~Vol.~14, No.~8, August~2015}%
{Shell \MakeLowercase{\textit{et al.}}: Bare Demo of IEEEtran.cls for Computer Society Journals}

\IEEEtitleabstractindextext{%
\begin{abstract}
Factorization machines (FMs) are widely used in recommender systems due to their adaptability and ability to learn from sparse data. However, for the ubiquitous non-interactive features in sparse data, existing FMs can only estimate the parameters corresponding to these features via the inner product of their embeddings. Undeniably, they cannot learn the direct interactions of these features, which limits the model's expressive power. To this end, we first present MixFM, inspired by Mixup, to generate auxiliary training data to boost FMs. Unlike existing augmentation strategies that require labor costs and expertise to collect additional information such as position and fields, these extra data generated by MixFM only by the convex combination of the raw ones without any professional knowledge support. More importantly, if the parent samples to be mixed have non-interactive features, MixFM will establish their direct interactions. Second, considering that MixFM may generate redundant or even detrimental instances, we further put forward a novel Factorization Machine powered by Saliency-guided Mixup (denoted as SMFM). Guided by the customized saliency, SMFM can generate more informative neighbor data. Through theoretical analysis, we prove that the proposed methods minimize the upper bound of the generalization error, which hold a beneficial effect on enhancing FMs. Significantly, we give the first generalization bound of FM, implying the generalization requires more data and a smaller embedding size under the sufficient representation capability. Finally, extensive experiments on five datasets confirm that our approaches are superior to baselines. Besides, the results show that "poisoning" mixed data is likewise beneficial to the FM variants.
\end{abstract}

}

\maketitle
\IEEEdisplaynontitleabstractindextext
\IEEEpeerreviewmaketitle

\IEEEraisesectionheading{\section{Introduction}\label{sec:introduction}}
\IEEEPARstart{T}{he}  rapid advancement of information technology has exacerbated information overload. As an effective way to alleviate information overload, recommender systems have attracted many researchers to study \cite{yang2016social,lian2020lightrec,nguyen2014gaussian,lian2018xdeepfm}. In the actual recommendation task, the sample's features are mostly categorical, so they become highly sparse and difficult to learn after one-hot encoding. For this reason, factorization machines (FMs) \cite{rendle2010factorization} are proposed. 
On the one hand, FMs can learn model parameters and task predictions in linear time, which is necessary for large-scale systems. On the other hand, FMs can effectively estimate highly sparse features' interactions and generalize to unobserved ones. The flexibility and effectiveness make them widely used in e-commerce \cite{chen2021automated}, location services \cite{oentaryo2014predicting}, and social networks \cite{xie2018factorization}.

Despite the pleasant expression capability for sparse data, the performance of FM is heavily dependent on real-world datasets with complex and nonlinear underlying structures \cite{he2017neural} (e.g., overfitting to the infrequently-occurring features or underfitting to the frequent features \cite{chen2019rafm}). Accordingly, it has remained challenging to perform high-quality predictions when facing data outside the learned discrete feature space. For this purpose, some work attempted to enhance the models by incorporating auxiliary information, e.g., Fields \cite{juan2016field}, time information \cite{rendle2011fast}, multi-view \cite{lu2017multilinear}, and position information \cite{wang2018contextual}. Apparently, such methods need to ensure that the additional information is available or easy to collect. Apart from the time-consuming labor costs, they also need the support of professional knowledge, which limits their applicability. Therefore, it is necessary to study efficient data augmentation without additional knowledge or manual intervention.

In addition, the key of FM is feature interaction, but it is pretty common for some feature interactions to be absent or not observed under limited sparse data. For instance, it is well known that smoking is forbidden in public, whereby the features corresponding to smoking and public places will not appear in moral users simultaneously. If an immoral user (who often smokes in public) appears, the model may not perform well for this user. Although FM attempts to estimate the coefficients between such non-interactive features via an inner product, it is irrefutable that the model cannot learn the direct interaction of these features. Constructing explicit interactions between such features may help learn from ignored users that are outside the training set, thereby enhancing the recommendation generalization.

Given the aforementioned deficiencies and inspired by the augmentation capability of Mixup \cite{zhang2017mixup} in computer vision, we first propose Mixup powered Factorization Machine (MixFM) to enhance FMs. It is the first attempt of Mixup at sparse recommendations, which can simultaneously consider the knowledge limitations and explicit interaction problem of non-interactive features. Specifically, through the raw samples' convex combination, MixFM generates many neighbor data to enrich the dataset without professional knowledge. These auxiliary data enable MixFM to process the regions between natural sparse samples in a continuous linear manner, thereby alleviating the dependence on raw data. Besides, if the parent samples to be mixed have non-interactive features, MixFM will inherit these features and establish their direct interaction. Second, seeing that the randomness of neighbor data generated by MixFM may produce meaningless data, we further develop a novel Saliency-guided Mixup powered Factorization Machine (SMFM). In SMFM, we provide a solver to efficiently calculate the sample's saliency guiding SMFM to create informative data. We theoretically prove that our proposals 
hold a smaller upper bound of generalization error than FM. Besides, we give the first generalization bound of FM and provably reveal the impact of embedding and data on recommendation generalization. Notably, this is the first to investigate the generalization properties.

The contributions are multi-fold:
\begin{itemize}
	\item We give the first generalization bound of factorization machines. The bound reveals that the generalization requires more data, and a smaller embedding size under the sufficient representation capability of FM, which sheds light on further enhancing FMs.
	\item We propose two novel approaches, MixFM and SMFM, to boost FMs by generating auxiliary data. They do not need additional knowledge and could establish direct interaction with original non-interactive features. Besides, we contribute an efficient solution for calculating the sample's saliency guiding SMFM to generate more informative data.
	\item We deliver the generalization bound of MixFM which incorporates auxiliary data, and theoretically prove that our methods have a better generalization guarantee than FM.
	\item Through extensive experiments with five real-world datasets, we empirically demonstrate the superiority of MixFM and SMFM in improving performance. Furthermore, the results indicate that our methods are also effective in FM variants.
\end{itemize}

\section{Related work}
Given that traditional linear models deal with features independently and ignore their interactions, Rendle et al. \cite{rendle2010factorization} proposed factorization machines (FMs), which use a simple dot product to achieve feature interaction. Unlike traditional Matrix Factorization \cite{jiang2019heterogeneous}, which only models the interactions between users and items, FM can arbitrarily combine various features. Such flexibility makes it widely used in various practical scenarios \cite{nguyen2014gaussian,li2019android,oentaryo2014predicting,yamada2017convex,lian2016regularized}. However, these specific features require expert knowledge support and expensive labor, which limit their applicability. This encourages us to investigate more general augmentation algorithms.

In recent years, many FM variants have been proposed and achieved promising performance in specific tasks. HOFM \cite{blondel2016higher} carried out third-order or higher-order extensions of feature interaction. Considering that FM treats the contribution of each pair feature interactions equally and ignores its differences, Juan et al. \cite{juan2016field} proposed Field-aware factorization machines (FFM). They introduced the concept of field and learned its embedding representation for each feature in each field. Obviously, the parameters learned by FFM are expanded by $F$ times, where $F$ is the field size. For this reason, AFM \cite{xiao2017attentional} was proposed, which uses the attention module to distinguish the interaction of different field features. Compared with FM, AFM's additional parameters are negligible. Inspired by adversarial training, Punjabi et al. \cite{punjabi2018robust} proposed a robust factorization machine by adding perturbations to the first-order and second-order terms during training. Cheng et al. \cite{cheng2014gradient} proposed a greedy interactive feature selection operator based on gradient boosting and combined it with FM. Recently, multi-view machines \cite{lu2017multilinear,lu2018learning,cao2016multi} that model high-order feature interactions based on multi-views have begun to receive attention, and they can design feature interactions from different perspectives. We cannot deny the massive memory footprint of the model when faced with hundreds to thousands of categorical features. Therefore, it is essential to reduce the memory footprint without significantly reducing the performance \cite{jiang2021xlightfm,liu2018discrete}. 

Benefiting from the potential of deep neural networks (DNN) in feature representation, some work \cite{shan2016deep,guo2017deepfm,he2017neural,wang2017deep,lian2018xdeepfm} utilized them to model high-order interactions. DeepFM \cite{guo2017deepfm} used a nonlinear DNN to capture high-order feature interactions and combined them with FM's low-order feature combinations. Since the stitching structure of DeepFM is challenging to characterize the second-order crossover features completely, NFM \cite{he2017neural} only used DNN after the linear second-order interaction term to enhance the expression of the second-order interactive information. Inspired by the Deep \& Cross network (DCN) \cite{wang2017deep} could automatically construct high-order iterative features, Lian et al. \cite{lian2018xdeepfm} proposed XDeepFM, which combines the vector-wise of FM with DCN. Although high-order interactions enhance the model representation, it is still difficult to directly learn unobserved feature interactions. This generalization gap leaves room from improvement in their recommendation performance.

\section{Preliminaries}
\subsection{Factorization Machines}
\begin{figure*}[h]
	\centering
	\includegraphics[width=0.96\linewidth]{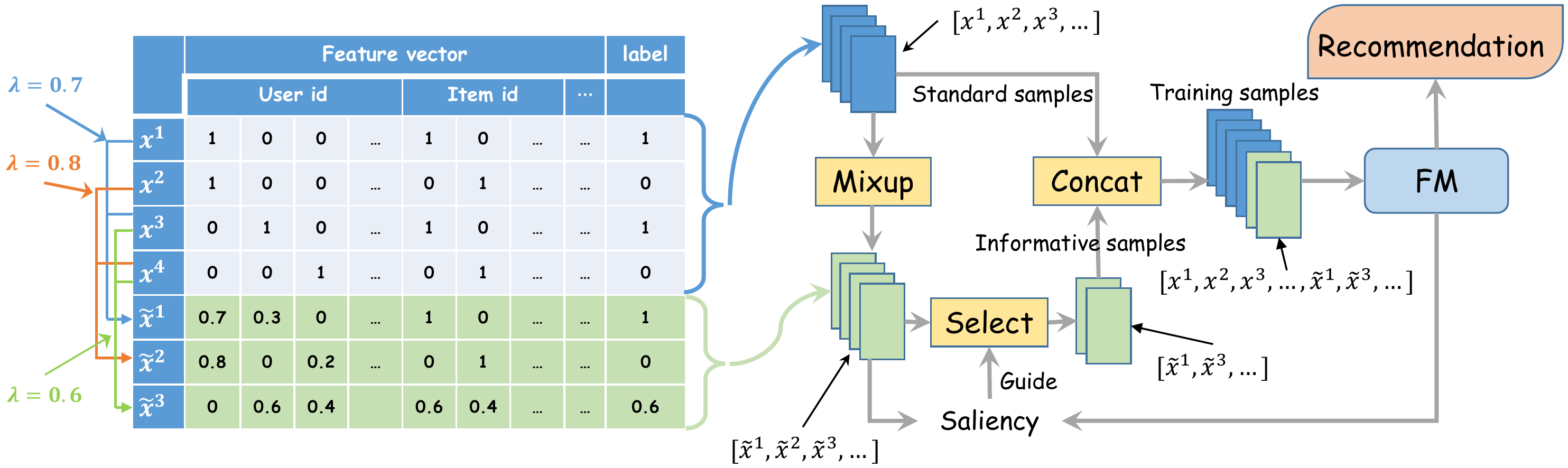}
	\caption{Overview of our methods. Left part: An example of mixed data in FM, where $\{x^1,\ldots,x^4\}$ represent the natural samples, and $\{\tilde{x}^1,\tilde{x}^2,\tilde{x}^3\}$ are the mixed samples. Right part: the framework of SMFM.}
	\label{fig: framework}
\end{figure*}
\label{subsec:fm}
In the practical recommendation, data can be described by categorical and numerical features. Below we give an example.
$$
 \overbrace{\underbrace{[1,0, \ldots, 0]}_{\text {User ID }} \underbrace{[0,1,\ldots,0]}_{\text {Item ID }} \underbrace{[1,0,1, \ldots]}_{\text {Historical Items }}}^{\text{Categorical Features}}\overbrace{[0.1,0.29, \ldots,0.01]}^{\text {Numerical Features}}
$$
For category features, User ID and Item ID use one-hot encoding, while the historical item list employs multi-hot encoding. Numerical features are normalized to $[0,1]$. Clearly, even if there are only three types of categorical features in this example, the feature vector will become highly sparse after encoding.

For this reason, Rendle et al. \cite{rendle2010factorization} proposed factorization machines for prediction tasks of extremely sparse data. Its key idea is to consider the interrelationships between sparse features and combine them. In the paper, we only study the 2-way interactive FM. Specifically, assuming a real-valued feature vector $x\in \mathcal{R}^m$ (most values are 0), the learning paradigm of FM can be defined as 
\begin{equation}
f(x)=w_{0}+\sum_{i=1}^{m} w_{i} x_{i}+\sum_{i=1}^{m} \sum_{j=i+1}^{m}\left\langle v_{i}, v_{j}\right\rangle x_{i} x_{j},
\label{eq: fm_origin}
\end{equation}
where $m$ is the number of features after encoding, $w_0$, $w_i$ are the weights of the bias term and the linear term, respectively. $v\in\mathcal{R}^{m\times d}$, and $v_i\in\mathcal{R}^d$ represents the embedding vector of the $i$-th feature $x_i$, where $d$ is the embedding size. The inner product $\langle v_{i},v_{j}\rangle$ parameterizes the interaction between $x_i$ and $x_j$. It is worth noting that the computational complexity of Eq. \ref{eq: fm_origin} is $\mathcal{O}(dm^2)$. To efficiently work, the second-order term is usually reformulated as
\begin{equation}
\begin{aligned}
\sum_{i=1}^{m} \sum_{j=i+1}^{m}\left\langle{v}_{i}, {v}_{j}\right\rangle x_{i} x_{j}
= \frac{1}{2} \sum_{k=1}^{d}\left[\left(\sum_{i=1}^{d} v_{i, k} x_{i}\right)^{2}-\sum_{i=1}^{d} v_{i, k}^{2} x_{i}^{2}\right].
\label{eq: second_order}
\end{aligned}
\end{equation}
The complexity is reduced to $\mathcal{O}(dm)$. 

\subsection{Saliency Map}
\label{sec: saliencymap}
With the rapid development of machine learning today, even in the face of a model with good performance, we will still be wary as the model is not interpretable, and we fail to know the working principle. Studying the interpretability may help clarify the model's internal mechanisms and working principles \cite{zhang2021group,xu2022towards,zhou2018interpreting}. One of the most widely studied tools is the saliency map \cite{simonyan2014deep}, which generates a saliency vector with the same size as input to determine which features dominate the decision. For example, for an image $x$ in the field of computer vision, $f(x)$ represents the prediction of $x$ by the model $f$, then the saliency map $\mathcal{S}$ of any image $x_0$ can be calculated by:
$$\mathcal{S}(x_0,f)=\left.\frac{\partial f(x)}{\partial x}\right |_{x_0},$$
where the i-th value of $\mathcal{S}(x_0,f)$ represents the saliency of the i-th pixel of $x_0$. 
\section{Methodology}
This section will detail the proposed boosting strategies: MixFM, and its improved version, SMFM.

\subsection{MixFM}

First, we present a domain-knowledge-independent method, namely Mixup powered Factorization Machine (MixFM), for enhancing the FMs. It is mainly based on the following considerations: (1) learning extremely sparse data will be highly dependent on (overfit or underfit) the dataset; (2) existing information assistance methods require expensive workforce collection or assurance of information availability; (3) FM and its variants cannot directly establish the explicit relationship between the non-interactive features that are not observed or even absent, which limits their performance to generalize to unseen data.

Unlike traditional FM, which performs empirical risk minimization (ERM) on the original dataset, MixFM optimizes based on the principle of vicinal risk minimization (VRM). Specifically, MixFM is trained on both the standard dataset and the neighbors of natural samples. Suppose $\mathcal{D}$ is the standard dataset. For any sample pair $(x, y)\sim \mathcal{D}$, its neighbor data $(\tilde{x},\tilde{y})$ is defined as follows:
\begin{equation}
	\begin{split}
	&\tilde{x}=\lambda\cdot x+(1-\lambda)\cdot r_x,\\
	&\tilde{y}=\lambda\cdot y+(1-\lambda)\cdot r_y,
	\end{split}
	\label{eq: mix}
\end{equation}
where $\lambda= \max (\lambda',1-\lambda')$, and $\lambda'\sim Beta(\alpha,\beta)$. $\alpha$ and $\beta$ are controllable parameters for Beta distribution, and $(r_x,r_y)$ is a data pair sampled from the dataset $\mathcal{D}$. Unlike Mixup, we limit $\lambda\ge 0.5$ (indicated by $\lambda=\max (\lambda',1-\lambda')$). In this way, we can customize the first parent sample $(x,y)$ to make the nearest neighbor of the generated data controllable, thereby avoiding the generated samples being biased towards specific ones and increasing the data diversity.

The left part of Fig. \ref{fig: framework} illustrates the process of generating new samples in MixFM. As can be seen, the generated mixed data inherits the features and behaviors of the two parents: (1) If the parent instances have the same behavior (feedback or not), then the mixed samples inherit the features of both and exhibit the same behavior, e.g., $\tilde{x}^1$ and $\tilde{x}^2$ in Fig. \ref{fig: framework}. (2) If the parents show different behaviors, then the mixed sample makes a compromise, e.g., $\tilde{x}^3$ in Fig. \ref{fig: framework}. Intuitively, 
such compromise of reducing weights avoids overconfident decisions for uncertain feature interactions.

\begin{definition}[MixFM]
	\label{def: mixfm}
	Suppose the original dataset is $\mathcal{D}=\{(x^i,y^i)\}_{i=1}^n$ of size $n$, $n'$ mixed samples $\widetilde{\mathcal{D}}=\{(\tilde{x}^i,\tilde{y}^i)\}_{i=1}^{n'}$ are generated according to Eq. \ref{eq: mix}. Let $\mathcal{L}(f(x),y)$ be the training loss with respect to $(x,y)$, where $f$ represents the standard FM, then MixFM performs empirical vicinal risk minimization on $\mathcal{D}\cup\widetilde{\mathcal{D}}$:
	\begin{equation}
		\arg \min \limits_{f}\frac{1}{n+n'}\sum\nolimits_{(x,y)\in\mathcal{D}\cup\widetilde{\mathcal{D}}}\mathcal{L}(f(x),y).
	\label{eq: loss}
	\end{equation}
\end{definition}

Definition \ref{def: mixfm} gives the formal definition of MixFM. The detailed process of MixFM is shown in Alg. \ref{alg: MixFM}. To make the mixed samples cover the entire data space as possible (become all users' neighbors), we perform sampling without replacing the first parent sample (Line 3 in Alg. \ref{alg: MixFM}). If for the special case where the number of generated samples exceeds the dataset size (i.e., $n'>n$), then $(n'\ mod\ n)$ data are sampled without replacement, and copy the entire dataset $\lfloor n'/n \rfloor$ times, where $\lfloor\cdot\rfloor$ is rounded down.

Reconsidering MixFM, firstly, MixFM can generate massive neighbor data, which may alleviate the model's high dependence on raw data. Secondly, the generated data is not supported by any professional knowledge. It is worth noting that we only need to mix non-zero features for the sparse data, and the complexity of generating the new sample is approximately $\mathcal{O}(1)$, which cannot be achieved in other fields such as computer vision. Finally, for any non-interactive feature pair, as long as they respectively exist in the parent samples, the mixed data can establish their direct interaction by inheriting all parent samples' features. For instance, in Fig. \ref{fig: framework}, the third feature component of user id and the first component of item id have no intersection in natural data $\{x^i\}_{i=1}^4$, while the mixed sample $\tilde{x}^3$ establishes their interaction. Collectively, MixFM could make up for the limitations of existing FMs.

\begin{algorithm}[tb]
	\caption{MixFM}
	\label{alg: MixFM}
	\textbf{Input:} 
	The dataset $\mathcal{D}=\{(x^i,y^i)\}_{i=1}^n$;
	training epochs $T$;
	the number of mixed data $n'$.
	\begin{algorithmic}[1]
		\FOR{$t=1\dots T$}
    	\STATE $\widetilde{\mathcal{D}}=\varnothing$.
    	\STATE Select $n'$ instances $\widehat{\mathcal{D}}$ from $\mathcal{D}$ without replacement.
    	\FOR{each $(x,y)$ in $\widehat{\mathcal{D}}$}
    	\STATE Generate $(x,y)$'s neighbor $(\tilde{x},\tilde{y})$ according to Eq. \ref{eq: mix}.
    	\STATE $\widetilde{\mathcal{D}}=\widetilde{\mathcal{D}}\cup\{(\tilde{x},\tilde{y})\}$.
    	\ENDFOR
    	\STATE Do standard training on $\mathcal{D}\cup\widetilde{\mathcal{D}}$ according to Eq. \ref{eq: loss}.
		\ENDFOR
		\RETURN $\mathcal{D}'$
	\end{algorithmic}
\end{algorithm}

\subsection{SMFM}

The randomness of the samples generated by MixFM makes it inevitable to produce some redundant or even detrimental samples. For instance, if the $\lambda$ is infinitely close to 1, MixFM is similar to copying raw samples, which only increases the training cost under sufficient training. Therefore, it is vital to generate more meaningful samples. On the basis of MixFM, we further propose a Saliency-guided Mixup powered Factorization Machine (SMFM).

For a well-trained model, if adding a mild noise to a sample can cause a significant change in the decision, then the sample is likely to be in the blind zone of the model. Given that MixFM tries to model the discrete data space in a continuous manner, the samples corresponding to such blind domain are likely to be informative samples, and incorporating them into the dataset may be beneficial for training. Formally, suppose $\eta$ is the noise, where $\left\|\eta\right\|<\epsilon$. Then, the change of loss $\mathcal{L}(f(x),y)$ by adding noise $\eta$ to $x$ can be described as:
$$\Delta\mathcal{L}(f(x),y)\overset{1st-order\ approx.}{\approx}\left(\frac{\partial \mathcal{L}(f(x),y)}{\partial x}\right)^T\cdot \eta,$$
where $\Delta\mathcal{L}(f(x),y)=\mathcal{L}(f(x+\eta),y)-\mathcal{L}(f(x),y)$.
The $i$-th component in $\frac{\partial \mathcal{L}(f(x),y)}{\partial x}$ measures the saliency of the $i$-th feature of $x$ in the model decision. Unlike the saliency map described in Section \ref{sec: saliencymap}, we measure the significance of the loss rather than that of the prediction score. In addition, considering that the smaller the feature in FM, the lower its contribution to the prediction (there is no contribution when the feature is 0). Therefore, we define the weighted saliency of any sample $x$ in the paper:
\begin{equation}
	Saliency(x,y,f)=\left(\frac{\partial \mathcal{L}(f(x),y)}{\partial x}\right)^T\cdot x=\frac{\partial \mathcal{L}(f(x),y)}{\partial f}\cdot f(x),
	\label{eq: saliency}
\end{equation}
where the second equals sign is established according to the chain rule, and the complexity of calculating saliency is $\mathcal{O}(dm)$.  

With the means of calculating the sample's saliency, we can use it to guide the generator to explore more significant neighbor data, which is the core idea of SMFM. Specifically, in SMFM, we first generate $p$ candidate neighbors for each selected sample by Eq. \ref{eq: mix}. Second, according to Eq. \ref{eq: saliency}, the most salient neighbor is selected from the $p$ neighbors as the final mixed sample. The framework of SMFM is illustrated in the right part of Fig. \ref{fig: framework}, and the detailed algorithm is shown in Alg. \ref{alg: SMFM}.
\begin{algorithm}[tb]
	\caption{SMFM}
	\label{alg: SMFM}
	\textbf{Input:} 
	The dataset $\mathcal{D}=\{(x^i,y^i)\}_{i=1}^n$ with size $n$;
	training epochs $T$;
	the size of mixed data $n'$;
	the number of candidate neighbors per sample $p$.
	\begin{algorithmic}[1]
		\FOR{$t=1\dots T$}
    	\STATE $\widetilde{\mathcal{D}}=\varnothing$.
    	\STATE Select $n'$ instances $\widehat{\mathcal{D}}$ from $\mathcal{D}$ without replacement.
    	\FOR{each $(x,y)$ in $\widehat{\mathcal{D}}$}
    	\STATE Generate $p$ neighbors of $(x,y)$ through Eq. \ref{eq: mix}, denote them as $\widehat{\mathcal{D}}_{(x,y)}$.
		\STATE $(\tilde{x},\tilde{y})=\arg\max\limits_{(x',y')\in\widehat{\mathcal{D}}_{(x,y)}}Saliency(x',y',f)$.
		\STATE $\widetilde{\mathcal{D}}=\widetilde{\mathcal{D}}\cup\{(\tilde{x},\tilde{y})\}$.
    	\ENDFOR
    	\STATE Do standard training on $\mathcal{D}\cup\widetilde{\mathcal{D}}$ according to Eq. \ref{eq: loss}.
		\ENDFOR
		\RETURN $\mathcal{D}'$
	\end{algorithmic}
\end{algorithm}

Theoretically, the proposed strategies could work as long as the model is first-order differentiable. However, due to models' diversity, broad applicability is hard to guarantee. For FMs, which are widely used in the critical re-ranking stage, we can strictly guarantee the effectiveness from empirical and theoretical aspects. Therefore, we mainly focus on FMs in the paper.

It is worth noting that the recent work \cite{gong2020maxup} proposed Maxup, but the proposed SMFM is quite different from them. On the one hand, the mixed data in Maxup is selected based on the loss, whereas we utilize the saliency of the sample. On the other hand, Maxup modifies the natural training data; in contrast, we "poison" new data without changing the original ones. We also perform detailed comparisons in the experiments.
\section{Theoretical Results}
\label{sec: theoretical}
This section theoretically analyzes the positive role of our methods in improving the generalization of FM. For the output hypothesis $f$ of the learning algorithm, the generalization error $R_{\mathcal{D}}(f)$ we focus on is the predictive ability on unseen data, which cannot be directly observed, but its empirical error $\hat{R}_{\mathcal{D}}(f)$ on the training set can be directly observed. To this end, we explore the generalization error upper bound concerning empirical error, which gives the worst generalization performance. 

Notably, although related work \cite{zhang2020does} has delivered the generalization bound of Mixup, it is limited to the typical linear model with input constraints and does not apply to FM with second-order or higher interactions. Nevertheless, we can benefit from their work by ignoring the linear term $\sum_{i=1}^{m}w_ix_i$ in FM for simplification. All proofs can be found in Section \ref{proofs}.

\subsection{Generalization Bound for Factorization Machines}
We give the necessary notations that will be used in proofs. $R(x,k)$ denotes repeating $x$ $k$ times (e.g., $[R(3,2),R(2,1)]=[3,3,2]$). Besides, we define $\bigvee_{i=1}^nx_i=x_1,\ldots,x_n$ (e.g., $[\bigvee_{i=1}^3x_i]=[x_1,x_2,x_3]$). Then we can reformulate FM as follows.

\begin{definition}[Linear representation of FM]
	\label{def: linear_fm}
	For an FM that omits the first-order term, define a feature vector $u=[\bigvee_{i=1}^{m}\bigvee_{j=i+1}^{m}R(x_ix_j,d), R(0,d^2m^2-dm(m-1)/2)]^T\\ \in\mathcal{R}^{d^2m^2}$, $\theta=[\bigvee_{i=1}^{m}\bigvee_{j=i+1}^{m}\bigvee_{k=1}^{d}v_{i,k}v_{j,k},\\\bigvee_{i=1}^{m}\bigvee_{j=i+1}^{m}\bigvee_{k=1}^{d}\bigvee_{l\ne k}v_{i,k}v_{j,l},\\ \bigvee_{i=1}^{m}\bigvee_{j=1}^{i}\bigvee_{k=1}^{d}\bigvee_{l=1}^{d}v_{i,k}v_{j,l}]^T\in\mathcal{R}^{d^2m^2}$, where $x$ and $v$ are consistent with the expression in Section \ref{subsec:fm}. Then, FM can be redefined as
	\begin{equation}
	\label{eq: linear_fm}
		f(u)=\theta^Tu+w_0.
	\end{equation}
\end{definition}

Comparing Eq \ref{eq: linear_fm} with Eq. \ref{eq: second_order}, it is easy to find that they are equivalent. Following \cite{arora2021dropout}, we study the following function class $\mathcal{H}_{\gamma}$ closely related to FM:
\begin{equation}
	\mathcal{H}_{\gamma}:=\{f:x\mapsto f(x), \sum_{i=1}^m\sum_{j=1}^{d}v_{i,j}^2\le\gamma\}.
\end{equation}
By Definition \ref{def: linear_fm} and $(\sum_{i=1}^m\sum_{j=1}^{d}v_{i,j}^2)^2=\left\|\theta\right\|^2$, this is equivalent to
\begin{equation}
\mathcal{H}_{\gamma}:=\{f:u\mapsto \theta^Tu+w_0, \left\|\theta\right\|^2\le\gamma^2\}.
\label{problem: fm}
\end{equation}

In classical learning theory, the model generalization is closely related to the function complexity, and the typical complexity measurements are VC-Dimension \cite{blumer1992learnability}, Covering number \cite{zhou2002covering}, and Rademacher complexity \cite{bartlett2002rademacher}. We adopt Rademacher complexity to analyze the complexity of FM in Eq. \ref{problem: fm}.
\begin{theorem}
	\label{theorem: rademacher_fm}
	For an FM that omits the first-order term, suppose its hypothesis space is $\mathcal{H}_{\gamma}$ and $\mathcal{D}=\{(x^i,y^i)\}_{i=1}^n$ is the dataset of size $n$, where $\left\|x^i\right\|_{\infty}\le 1$, and $\left\|x^i\right\|_0\le\tau$. Then the empirical Rademacher complexity $\widehat{\Re}_{\mathcal{D}}(\mathcal{H}_{\gamma})$ satisfies
	\begin{equation}
		\widehat{\Re}_{\mathcal{D}}(\mathcal{H}_{\gamma})\le \sqrt{\frac{\gamma^2d\tau(\tau-1)}{2n}}.
		\nonumber
	\end{equation}
\end{theorem}
This theorem gives the empirical Rademacher complexity of FM. Furthermore, we derive the generalization bound of FM as follows.
\begin{corollary}
	Suppose that for the function $f$ in $\mathcal{H}_{\gamma}$, $f$ maps any sample to $[0,1]$. Sample $x$ satisfies $\left\|x\right\|_{\infty}\le 1$, and $\left\|x\right\|_0\le\tau$. Then, for any $\delta>0$, with probability at least $1-\delta$ over the draw of an $i.i.d.$ dataset $\mathcal{D}$ of size $n$, the following holds for all $f\in\mathcal{H}_{\gamma}$:
	\begin{equation}
		R_{\mathcal{D}}(f)\le \hat{R}_{\mathcal{D}}(f)+2\sqrt{\frac{\gamma^2d\tau(\tau-1)}{2n}}+3\sqrt{\frac{\ln (2/\delta)}{2n}},
		\label{eq: generalization_fm}
	\end{equation}
	where $R_{\mathcal{D}}(f)$ and $\hat{R}_{\mathcal{D}}(f)$ represent the generalization error and empirical error of the model, respectively.
	\label{theorem: generalization_fm}
\end{corollary}

\noindent\textbf{Remark 1.} Eq. \ref{eq: generalization_fm} indicates the FM's generalization error is related to the $\mathcal{O}(\sqrt{\frac{d\tau^2}{n}})$ term. On the one hand, a smaller embedding size $d$, more data $n$ will benefit model generalization. On the other hand, the generalization error is related to the empirical error $\hat{R}_{\mathcal{D}}(f)$. A smaller $d$ reduces model complexity, which in turn may increase empirical error. In summary, the model's generalization needs more data, but the embedding size $d$ requires a compromise between model complexity and empirical error. Significantly, this is the first generalization bound of FM, lacking in the previous work.

\subsection{Generalization Bound for MixFM}
\label{sec: mixfm_generalization}
After giving the generalization bound of the standard FM, we prove that MixFM has a smaller generalization error. In MixFM, the operation of injecting mixed samples $\widetilde{\mathcal{D}}$ into the dataset can be decomposed into two stages: (1) inject all first parent samples of $\widetilde{\mathcal{D}}$ and define them as $\widehat{\mathcal{D}}$; (2) Modify $\widehat{\mathcal{D}}$ to its neighbor data $\widetilde{\mathcal{D}}$ by Eq. \ref{eq: mix}. From the mathematical expectation, stage 1 increases the training epochs, so we have $R_{\mathcal{D}\cup\widehat{\mathcal{D}}}(f)=R_{\mathcal{D}}(f)$ under sufficient training (it will also be evidenced in the experiments). Therefore, we just need to analyze stage 2; that is, as long as performing Mixup on the original dataset provably reduces the generalization error, it will be concluded that the effectiveness of MixFM in boosting generalization. In the following analysis, we adopt logistic loss for the classification tasks. Note that the following analysis is also applicable to regression tasks and even has a more concise form.
\begin{theorem}
	For an FM that omits the first-order term, $\mathcal{D}=\{(x^i,y^i)\}_{i=1}^n$ is the dataset of size $n$, and $\widetilde{\mathcal{D}}=\{(\tilde{x}^i,\tilde{y}^i)\}_{i=1}^n$ is the dataset after Mixup, where $(\tilde{x}^i,\tilde{y}^i)$ is calculated by Eq. \ref{eq: mix}. Besides, $u$ and $\theta$ are defined in Definition \ref{def: linear_fm}. Let $\mathcal{L}$ be the logistic loss, then the effect of Mixup in FM can be regarded as a data-related regularized term, that is,
	\begin{equation}
    \begin{split}
		&\mathcal{L}({\widetilde{\mathcal{D}}})-\mathcal{L}({{\mathcal{D}}})\\
		=&\frac{1}{n}\sum_{i=1}^{n}g(f(u^i))(1-g(f(u^i)))\mathbb{E}_{\tilde{\lambda},r_x}\frac{(1-\tilde{\lambda})^4}{\tilde{\lambda}^4}\theta^T\hat{\Sigma}_u\theta.
		\label{eq: fm_reg}
	\end{split}
	\end{equation}
	Here $g(\cdot)$ is a sigmoid function, $f(u^i)=\theta^Tu^i+w_0$, and $\hat{\Sigma}_u=\frac{1}{n}\sum_{i=1}^{n}u^i(u^i)^T$, where $u^i$ is transformed from $x^i$ by Definition \ref{def: linear_fm}. $\tilde{\lambda}\sim\frac{\alpha}{\alpha+\beta}Beta(\alpha+1,\beta)+\frac{\beta}{\alpha+\beta}Beta(\beta+1,\alpha)$.
	\label{theorem: regularization}
\end{theorem}
Theorem \ref{theorem: regularization} suggests that using Mixup on the original dataset is equivalent to adding a regularization term for standard training. Adhering to the similar approach in \cite{rendle2010factorization}, we consider the following function class closely related to the dual problem of Theorem $\ref{theorem: regularization}$:
\begin{equation}
\widetilde{\mathcal{H}}_{\gamma}:=\{f:u\mapsto \theta^Tu+w_0, \mathbb{E}_ug(f(u))(1-g(f(u)))\theta^T{\Sigma}_u\theta\le\gamma\},
\label{problem: mixfm}
\end{equation}
where $f(u)=\theta^Tu+w_0$, ${\Sigma}_u=\mathbb{E}_u(uu^T)$. Next, we provide the generalization error bound of FM trained by mixed data.
\begin{theorem}
	Suppose that for the function $f$ in $\widetilde{\mathcal{H}}_{\gamma}$, $f$ maps any sample to $[0,1]$. the sample $x$ satisfies $\left\|x\right\|_{\infty}\le 1$, and $\left\|x\right\|_0\le\tau$. Then, for any $\delta>0$, with probability at least $1-\delta$ over the draw of an $i.i.d.$ dataset $\mathcal{D}$ of size $n$, the following holds for all $f\in\widetilde{\mathcal{H}}_{\gamma}$:
	$$\widetilde{R}_{\mathcal{D}}(f)\le \hat{R}_{\mathcal{D}}(f)+2\sqrt{\frac{(1+e)^2\gamma\cdot \tau(\tau-1)}{2en}}+3\sqrt{\frac{\ln (2/\delta)}{2n}},$$ 
	where $\widetilde{R}_{\mathcal{D}}(f)$ and $\hat{R}_{\mathcal{D}}(f)$ represent the generalization error and empirical error, respectively.
	\label{theorem: generalization_mixfm}
\end{theorem}
\noindent\textbf{Remark 2.} Comparing Corollary \ref{theorem: generalization_fm} and Theorem \ref{theorem: generalization_mixfm}, it is easy to conclude that when $\gamma\ge\frac{(1+e)^2}{ed}\approx\frac{5.08}{d}$, MixFM has a smaller generalization error bound. In the worst case, $\gamma\ge 2.54$ when $d=2$, which is still easily met because $\gamma$ is often large to enhance the expressive capability of the model. In our experiments, $\gamma$ takes the minimum value of 106.1 in the Frappe dataset, while in larger datasets, they are much larger than this value. Therefore, the expected risk of MixFM is closer to the empirical prediction, and the generalization ability is stronger \cite{mohri2018foundations}.

Inspired by \cite{papernot2016limitations}, saliency data can be regarded as adversarial examples. Therefore, SMFM that creates saliency neighbor samples for training essentially performs adversarial training on MixFM. Recent studies \cite{xing2021generalization,song2018improving} have proved that adversarial training helps to improve generalization, so the generalization of SMFM is not lower than MixFM, let alone FM. Thus far, we prove the effectiveness of the proposed methods in enhancing FM's generalization.
\section{Experiments}
\subsection{Experiments Settings}
\subsubsection{Datasets}
\begin{table}[htbp]
	\renewcommand\arraystretch{1.05}
	\centering
	\caption{Statistics of datasets}
	\begin{tabular}{c|cccc}
		\hline\noalign{\smallskip}
		Dataset & Instance & Feature & User  & Item \\
		\noalign{\smallskip}\hline\noalign{\smallskip}
		Frappe & 288,609  & 5,382  & 957   & 4,082  \\
		MovieLens & 2,006,859  & 90,445  & 17,045  & 23,743  \\
		Pet   & 1,235,316  & 844,273  & 740,984  & 103,287  \\
		Electronics & 7,824,481  & 4,677,697  & 4,201,695  & 476,000  \\
		Books & 22,507,155  & 10,356,390  & 8,026,323  & 2,330,065  \\
		\noalign{\smallskip}\hline
	\end{tabular}%
\label{tab:datasets}%
\end{table}%
We use five real-word datasets to validate the effectiveness of our methods. Table \ref{tab:datasets} lists the detailed statistics.

\textbf{Frappe}\cite{baltrunas2015frappe}\footnote{\url{https://www.baltrunas.info/context-aware}}: This is a context-aware mobile app dataset containing 288,609 records of 957 users on 4,082 apps. Each record information includes contextual information such as weather, time, and city. We use one-hot encoding and multi-hot encoding to encode each record into a 5,382-dimensional vector.

\textbf{MovieLens}\cite{harper2015movielens}\footnote{\url{http://grouplens.org/datasets/movielens/latest}}: This is a movie rating dataset, which includes 2,006,859 ratings of 23,743 movies by 17,045 users, and 668,953 movie tags are assigned to these movies. One-hot encoding is adopted to decompose each rating into 90,445 features.

\textbf{Pet Supplies (Pet), Electronics, Books}\footnote{\url{http://jmcauley.ucsd.edu/data/amazon/}}: They are composed of the review information collected by Amazon customers in three different domains, namely pet supplies, electronic products, and books. These datasets include 11,400,788 users' reviews on 103,287 pet items, 476,000 electronics, and 2,330,065 books. each review has the user ID, produce ID, user status, brand, and other information.

Since these datasets only contain positive instances, similar to \cite{he2017neural}, we randomly sample two negative instances paired with one positive instance. The augmented dataset is randomly divided into the training set, validation set, and test set with a ratio of 8:1:1.

\subsubsection{Comparison Methods}
\begin{table*}[htbp]
	\renewcommand\arraystretch{1.0}
	\centering
	\caption{Performance comparison. $*$, $**$, $***$ represent the improvement of our methods (bold) over base model (FM, NFM, AFM, DCN) are statistically significant for $p<0.05$, $p<0.01$, $p<0.001$, respectively.}
	\setlength{\tabcolsep}{0.0081\linewidth}{
		\begin{tabular}{c|c|c|c|c|c|c|c|c|c|c|c}
			\hline\noalign{\smallskip}
			\multirow{2}[0]{*}{} & \multirow{2}[0]{*}{Model} & \multicolumn{2}{c|}{Frappe} & \multicolumn{2}{c|}{MovieLens} & \multicolumn{2}{c|}{Pet} & \multicolumn{2}{c|}{Electronics} & \multicolumn{2}{c}{Books} \\
			\noalign{\smallskip}\cline{3-12}\noalign{\smallskip}
			&      & AUC $\uparrow$   & LogLoss $\downarrow$ & AUC $\uparrow$  & LogLoss $\downarrow$ & AUC $\uparrow$   & LogLoss $\downarrow$ & AUC $\uparrow$   & LogLoss $\downarrow$ & AUC $\uparrow$   & LogLoss $\downarrow$ \\
			\noalign{\smallskip}\hline\noalign{\smallskip}
			\multirow{18}[0]{*}{ \makecell[c]{Augmented\\ FM-based\\ models} } & FM    & 0.9653  & 0.2047  & 0.9446  & 0.2673  & 0.6776  & 0.4301  & 0.7100  & 0.4330  & 0.7634  & 0.2757  \\
			& CopyFM & 0.9705  & 0.1948  & 0.9454  & 0.2665  & 0.6739  & 0.4187  & 0.7099  & 0.4300  & 0.7621  & 0.2753  \\
			& \textbf{MixFM} & 0.9784  & \textbf{0.1537 } & 0.9508  & 0.2482  & 0.6846  & \textbf{0.4102 } & 0.7209  & \textbf{0.4233 } & 0.7684  & \textbf{0.2714 } \\
			& \textbf{SMFM}  & \textbf{0.9825 } & 0.1668  & \textbf{0.9558 } & \textbf{0.2381 } & \textbf{0.7023 } & 0.4136  & \textbf{0.7249 } & 0.4378  & \textbf{0.7759 } & 0.2810  \\
			\noalign{\smallskip}\cline{2-12}\noalign{\smallskip}
			& t-test & ***   & ***   & ***   & ***   & ***   & ***   & ***   & ***   & ***   & *** \\
			\noalign{\smallskip}\cline{2-12}\noalign{\vskip\doublerulesep\vskip-\arrayrulewidth}\cline{2-12}\noalign{\smallskip}
			& NFM   & 0.9696  & 0.2093  & 0.9403  & 0.3003  & 0.6718  & 0.4545  & 0.7171  & 0.4295  & 0.7736  & 0.2683  \\
			& CopyNFM & 0.9706  & 0.2088  & 0.9436  & 0.2819  & 0.6818  & 0.4260  & 0.7107  & 0.4361  & 0.7743  & 0.2667  \\
			& \textbf{MixNFM} & 0.9788  & \textbf{0.1404 } & 0.9417  & 0.2751  & 0.6877  & \textbf{0.4187 } & 0.7212  & \textbf{0.4244 } & 0.7756  & \textbf{0.2644 } \\
			& \textbf{SMNFM} & \textbf{0.9808 } & 0.1428  & \textbf{0.9529 } & \textbf{0.2697 } & \textbf{0.7075 } & 0.4376  & \textbf{0.7300 } & 0.4412  & \textbf{0.7766 } & 0.2845  \\
			\noalign{\smallskip}\cline{2-12}\noalign{\smallskip}
			& t-test & ***   & ***   & *     &       & ***   & **    & ***   &       & *     &  \\
			\noalign{\smallskip}\cline{2-12}\noalign{\vskip\doublerulesep\vskip-\arrayrulewidth}\cline{2-12}\noalign{\smallskip}
			& AFM   & 0.9651 & 0.2056 & 0.9447 & 0.2669 & 0.6778 & 0.4300  & 0.7100  & 0.4326 & 0.7634 & 0.2752 \\
			&CopyAFM & 0.9742 & 0.1920 & 0.9458 & 0.2684 & 0.6711 & 0.4273 & 0.7091 & 0.4303 & 0.7618 & 0.2754 \\
			& \textbf{MixAFM} & 0.9784 & \textbf{0.1536} & 0.9510 & 0.2475  & 0.6860 & \textbf{0.4091} & 0.7209 & \textbf{0.4232} & 0.7684 & \textbf{0.2684} \\
			& \textbf{SMAFM} & \textbf{0.9806} & 0.1612 & \textbf{0.9557} & \textbf{0.2372} & \textbf{0.6938} & 0.4120 & \textbf{0.7215} & 0.4363 & \textbf{0.7745} & 0.2801 \\
			\noalign{\smallskip}\cline{2-12}\noalign{\smallskip}
			& t-test & ***   & ***   & ***   & ***   & ***   & ***   & ***   & **    & ***   & * \\
			\noalign{\smallskip}\cline{2-12}\noalign{\vskip\doublerulesep\vskip-\arrayrulewidth}\cline{2-12}\noalign{\smallskip}
			& DCN   & 0.9753 & 0.2070 & 0.9505 & 0.2858 & 0.7077 & 0.3970 & 0.7197 & 0.4570 & 0.7626 & 0.2714 \\
			&CopyDCN & 0.9756 & 0.2223 & 0.9484 & 0.2919 & 0.6942 & 0.4096 & 0.7200  & 0.4219 & 0.7607 & 0.2773 \\
			& \textbf{MixDCN} & 0.9767 & \textbf{0.1811} & 0.9532 & \textbf{0.2646} & 0.7093 & 0.3962 & \textbf{0.7239} & \textbf{0.4310} & 0.7677 & 0.2685 \\
			& \textbf{SMDCN} & \textbf{0.9792} & 0.1857 & \textbf{0.9553} & 0.2808 & \textbf{0.7110} & \textbf{0.3952} & 0.7230 & 0.4327 & \textbf{0.7682} & \textbf{0.2676} \\
			\noalign{\smallskip}\cline{2-12}\noalign{\smallskip}
			& t-test &       & ***   & *     & ***   &       &       & *     & ***   & *     &  \\
			\noalign{\smallskip}\hline\noalign{\smallskip}
			\multirow{6}[0]{*}{\makecell[c]{Other\\ models}} & DeepFM & 0.9753 & 0.2114 & 0.9361 & 0.2883 & 0.6988 & 0.4039 & 0.7289 & 0.4128 & 0.7735 & 0.2656 \\
			& MaxFM & 0.9802 & 0.3374 & 0.9544 & 0.3533 & 0.7094 & 0.6539 & 0.7122 & 0.6284 & 0.7685 & 0.5211 \\
			& MaxNFM & 0.9773 & 0.1656 & 0.9522 & 0.2490 & 0.7085 & 0.4179 & 0.7162 & 0.4669 & 0.7535 & 0.3013 \\
			& LightGCN & 0.9422 & 0.3718 & 0.5110 & 0.6469 & 0.5851 & 0.5691 & 0.6096 & 0.5571 & 0.5731 & 0.6710 \\
			& DAE   & 0.8352 & 0.3317 & 0.5060 & 0.6897 & 0.5735 & 0.6901 & 0.5655 & 0.6891 & -      &-  \\
			& Mult-VAE & 0.8359 & 0.3297 & 0.5007 & 0.6861 & 0.5801 & 0.6823 & 0.5631 & 0.6910 & -      & - \\
			\noalign{\smallskip}\hline
		\end{tabular}%
	}
	\label{tab: comparison}%
\end{table*}%
The proposed strategies are compared to the following state-of-the-art feature interaction models and non-FM models.

\textbf{FM} \cite{rendle2010factorization}: This is the benchmark model, factorization machines, using pairwise feature interactions for prediction tasks.

\textbf{NFM} \cite{he2017neural}: Given that the FM with second-order feature interaction essentially works linearly, NFM introduces DNN and employs a Bi-Interaction Layer to process the second-order intersection information and strengthen its expressive ability.

\textbf{AFM} \cite{xiao2017attentional}: It introduces an attention module to reflect the model's different attention to different feature interactions.

\textbf{DCN (Deep \& Cross)} \cite{wang2017deep}: The cross layer automatically constructs high-order feature interactions with high efficiency, and the deep layer uses DNNs to explore the nonlinear feature interactions.

\textbf{DeepFM} \cite{guo2017deepfm}: It contains an FM module and a DNN module, which are used to extract low-order interactive features and high-order ones, respectively.

\textbf{CopyFM, CopyNFM, CopyAFM, CopyDCN}: They use the same FM-based models (FM, NFM, AFM, and DCN) but randomly double some raw samples. Since our strategies enrich the dataset, they are introduced to make a fair comparison.

\textbf{MixNFM, SMNFM, MixAFM, SMAFM, MixDCN, SMDCN}: 
It can be seen from Alg. \ref{alg: MixFM} and Alg. \ref{alg: SMFM} that the two proposed ideas are decoupled from factorization machines. Therefore, applying strategies similar to MixFM and SMFM, we introduce these models to validate the effectiveness of our ideas on FM variants. 

\textbf{MaxFM, MaxNFM}: They respectively represent that Maxup \cite{gong2020maxup}, a variant of Mixup, is used in FM and NFM.

\textbf{LightGCN} \cite{he2020lightgcn}: It treats the user-item interaction as a bipartite graph and uses a graph convolutional network customized for collaborative filtering for prediction.

\textbf{Mult-VAE, Mult-DAE} \cite{liang2018variational}: Mult-VAE is an extension of variational autoencoder in collaborative filtering, in which Mult-DAE is the denoising version. Note that they are not suitable for Books dataset with millions of items.

\subsubsection{Evaluation Metrics}
Given that the FM-based models are widely applied in the re-ranking stage \cite{lian2018xdeepfm,cheng2020adaptive}, we adopt two metrics commonly used in this stage, LogLoss (logistic loss) and AUC (Area under the ROC Curve), which are also used in existing FM-based models \cite{rendle2010factorization,rendle2011fast, blondel2016higher,juan2016field,lian2018xdeepfm}.
LogLoss measures the fitting degree between the estimated distribution and the real one, and AUC can well reflect the model performance even with unbalanced data. The smaller LogLoss indicates better performance, while AUC is the opposite. We do 30 independent repeated experiments and reported the average results. Besides, the paired t-test is performed to judge the statistical significance.
\subsubsection{Parameter Settings}
For all models, we use Adam to optimize logistic loss. The learning rate is selected from $[0.0001,0.0002,\ldots,0.001,0.005,0.01]$, the batch size is searched from $[128,256,512,\ldots,16384]$, and the embedding size is tuned from $[2,4,8,\ldots,64,128]$. Concretely, in Frappe and MovieLens, we set the embedding size to 64 and set it to 2 for the three datasets from Amazon. In the experiments, we notice that the larger the dataset, the smaller the learning rate is needed to ensure training stability. Therefore, related to the dataset size, we set the learning rate of the five datasets to $[0.001,0.001,0.001,0.0002,0.0002]$. To balance training efficiency and convergence rate, the batch size is set to 2048 in the first two datasets, and the last three ones are set to 16384. Unless otherwise specified, we set the candidate neighbors $p$ to 10, and the number of the mixed data $n'$ is the same as the raw dataset size. The source codes are available at GitHub: \url{https://github.com/Daftstone/SMFM}.
\subsection{Performance Comparison}

\begin{figure*}[htbp]
	\centering
	\includegraphics[width=0.96\linewidth]{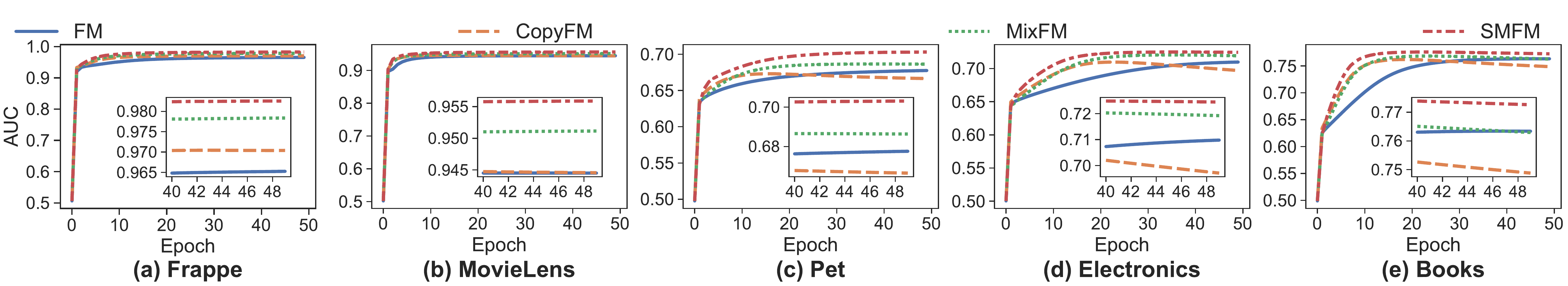}
	\caption{Test curves (AUC) for 50 epochs of training. The subplot in each figure shows the test curve of 40 to 50 epochs.}
	\label{fig: training_auc}
\end{figure*}
\begin{figure*}[htbp]
	\centering
	\includegraphics[width=0.96\linewidth]{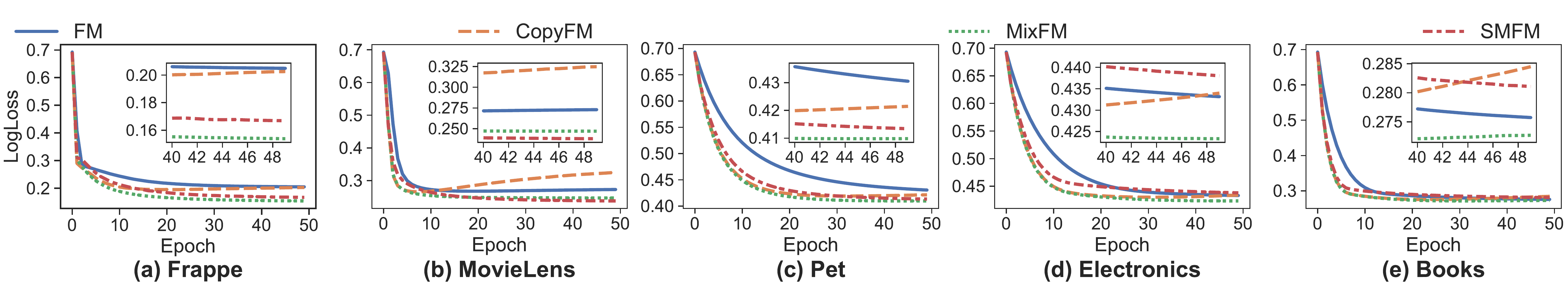}
	\caption{Test curves (LogLoss) for 50 epochs of training. The subplot in each figure shows the test curve of 40 to 50 epochs.}
	\label{fig: training_logloss}
\end{figure*}
We evaluate the performance in five datasets concerning AUC and LogLoss, and report the overall results in Table \ref{tab: comparison}. Noting that even 0.1 basis-points improvements are considered significant in CTR tasks because millions of impressions will magnify it daily. 

\subsubsection{Performance of Augmented FM-based Models}
The first four rows of results in Table \ref{tab: comparison} focus on the standard FM, and we have the following findings. First, for CopyFM, despite the doubled training samples, compared with FM, the performance is only a slight fluctuation, and even drops in some datasets (e.g., AUC in Pet). This finding aligns with the analysis in Section \ref{sec: mixfm_generalization} that copying samples cannot improve model generalization under sufficient training. Second, MixFM has a significant improvement compared to FM. The performance has an average improvement of 9.5\% for LogLoss, and in Frappe, the improvement even reaches 33.2\%. The gratifying result echoes Theorem \ref{theorem: generalization_mixfm} that MixFM has a smaller generalization error. Finally, SMFM achieves the best performance among all methods with respect to AUC, which implies the significant effect of using informative data. However, its LogLoss is not as good as MixFM, and in Books, it is even comparatively the worst. We reasonably suspect that SMFM essentially performs adversarial training on MixFM (in line with the analysis in Section \ref{sec: mixfm_generalization}), while adversarial training is often difficult to enjoy both prediction accuracy and model loss \cite{liu2020loss}. Given the above findings, if the task requirements tend to prediction accuracy, we suggest using SMFM, otherwise MixFM.

Furthermore, we illustrate the test curve of AUC and LogLoss during training, as shown in Fig. \ref{fig: training_auc} and Fig. \ref{fig: training_logloss}. Consistent with the results in Table \ref{tab: comparison}, our methods are in the lead. Besides, at the end of the training, overfitting appears in CopyFM but does not appear in MixFM and SMFM using the same amount of data. We reasonably believe that our methods add a data-adaptive regularization term, as shown in Theorem \ref{theorem: regularization}, which helps to alleviate overfitting.

We also migrate the proposed strategies to NFM, AFM, and DCN, to validate the effectiveness on FM variants, as shown in the middle part of Table \ref{tab: comparison}. These models configured with our strategies (denoted as MixNFM, SMNFM, MixAFM, SMAFM, MixDCN, SMDCN) still outperform the baselines, and similar findings to MixFM and SMFM can also be obtained. It demonstrates the potential of our methods in feature interaction models.

\subsubsection{Comparison with Other Models}
The lower part of Table \ref{tab: comparison} shows the performance of other FM variants and non-FM models. Firstly, as a variant of Mixup, although Maxup has some results close to our methods regarding AUC, the LogLoss is significantly inferior. This is because Maxup essentially performs adversarial training with a min-max optimization, challenging to enjoy both prediction performance and loss. Secondly, non-FM models such as LightGCN cannot use additional interactive features, resulting in poor performance in our experiments. Especially in MovieLens, where more than half of users do not have positive user-item interaction. Moreover, FM-based models studied in the paper are widely used in the re-ranking stage (metrics: AUC, LogLoss), while non-FM models such as Mul-VAE mainly focus on the matching stage (metrics: Recall, NDCG). Although the different emphases make a fair comparison impossible, the display of these results reflects the advantage of FM-based models in the re-ranking stage.

\subsection{Performance under Different Mixed Data $n'$}
\begin{figure}[htbp]
	\centering
	\includegraphics[width=1.\linewidth]{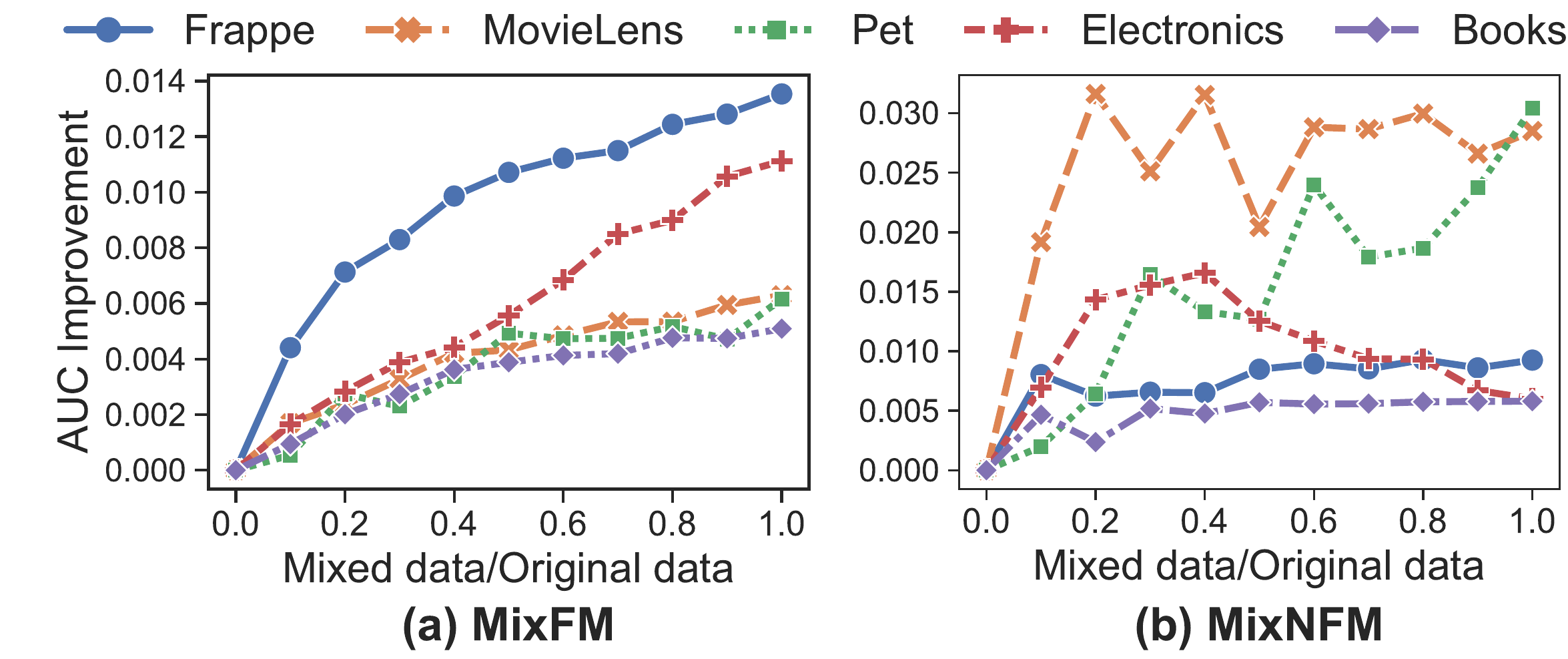}
	\caption{Performance under the different number of mixed samples $n'$.}
	\label{fig: ratio}
\end{figure}
In this section, we explore the effects of "poisoning" different numbers of mixed users ($n'$ in Alg. \ref{alg: MixFM}) on performance, and Fig. \ref{fig: ratio} reports the experimental results in MixFM and MixNFM. The abscissa represents the ratio of the injected neighbor data to the original sample; thus, 0 in abscissa means a standard FM. The ordinate denotes the improvement of AUC over standard FM. Overall, the model performance improves with the increase of mixed data. However, the improvement is not infinite, and the additional information seems to be saturated when the mixed data reaches a threshold, manifested in performance fluctuations (e.g., MixNFM on MovieLens) or decline (e.g., MixNFM on Electronics). In addition, the performance improvement of MixNFM is more significant than that of MixFM. We reasonably believe that NFM uses DNN to capture high-level interactions between features, which can mine the auxiliary information of the mixed data better than FM. Comparing different datasets, we find that the performance improvement will decrease as the data increases. Despite this, MixFM still has an AUC improvement of at least 0.004 when the data is doubled.
\subsection{Sensitivity w.r.t. the Number of Candidate Neighbors $p$}
\begin{figure}[htbp]
	\centering
	\includegraphics[width=1.\linewidth]{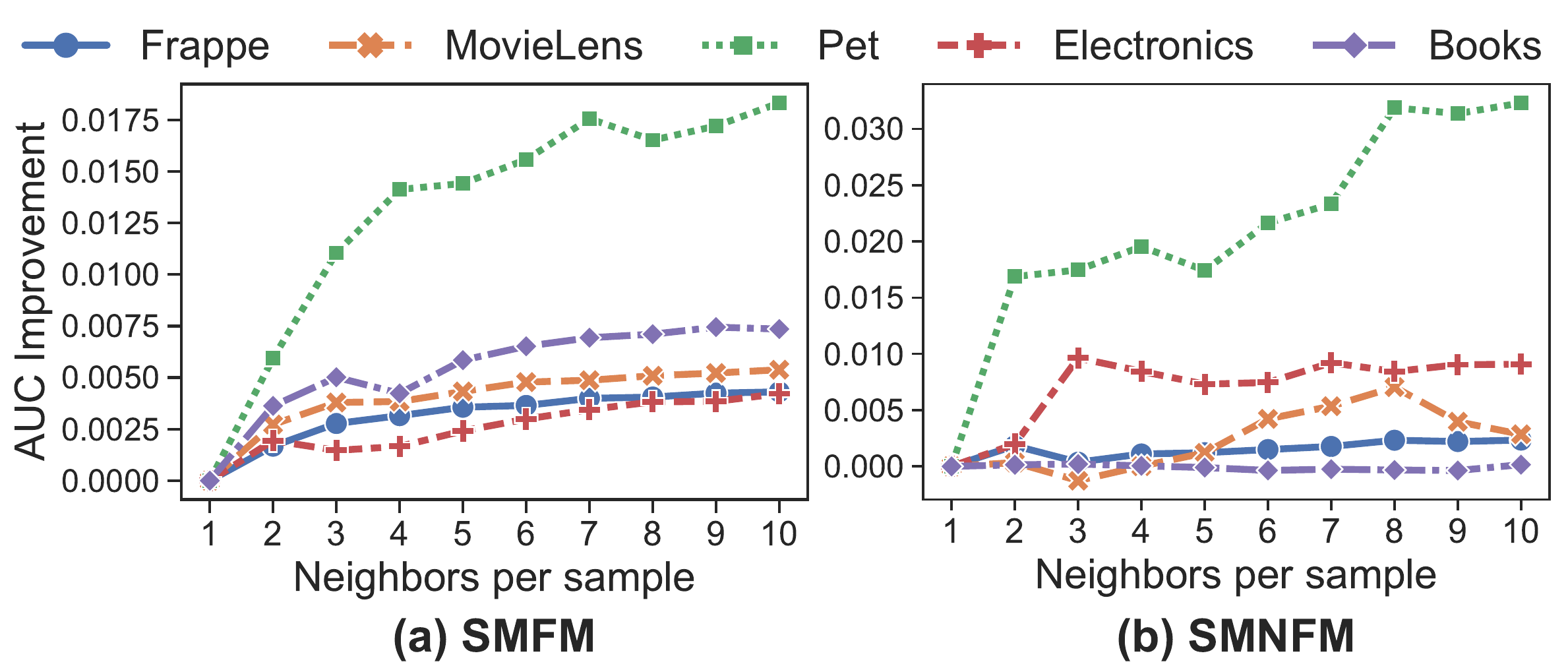}
	\caption{Performance under different neighbors $p$.}
	\label{fig: advratio}
\end{figure}
This section studies the performance sensitivity when using different numbers of candidate neighbors (p in Alg. \ref{alg: SMFM}) in SMFM and SMNFM, as shown in Fig. \ref{fig: advratio}. The abscissa denotes the number of candidate neighbors generated for each raw sample, and the ordinate is the performance improvement compared to MixFM and MixNFM, respectively. First, compared to generating one neighbor (MixFM or MixNFM), the performance of generating multiple neighbors to select the most significant one is improved, which stresses the effectiveness of the proposed Saliency-guided Mixup. Second, the sample's saliency (i.e., the number of candidate neighbors) and performance are not entirely positively correlated. For example, in MovieLens, when the number of neighbors generated is 8, the performance of SMNFM reaches saturation, and more neighbors even reduce the performance. This is mainly because excessively strong saliency samples will cause serious crossover mixture problems between raw and mixed data, thereby increasing the difficulty of training. Notably, a similar conclusion can be found in \cite{zhang2020attacks}.

\subsection{Sensitivity w.r.t. the Embedding Size $d$}
\begin{figure*}[htbp]
	\centering
	\includegraphics[width=1.\linewidth]{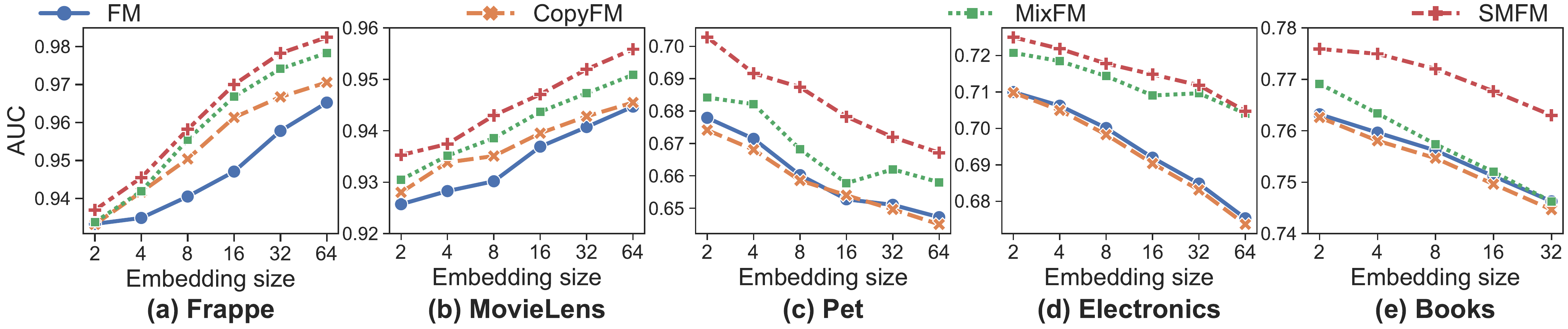}
	\caption{Performance comparison (AUC) under different embedding sizes $d$.}
	\label{fig: embedding_auc}
\end{figure*}
The embedded size $d$ plays a vital role in capturing the features' interaction. Besides, we analyze that the embedding size needs to compromise between model complexity and empirical error in Corollary \ref{theorem: generalization_fm}. The performance comparison concerning AUC under different embedding sizes is illustrated in Fig. \ref{fig: embedding_auc}.

Firstly, MixFM and SMFM under various settings are still leading, and SMFM achieves the best performance in all experimental models. The result emphasizes the effectiveness of our methods. Secondly, in Frappe and MovieLens, the performance positively correlates with the embedding size, but it is the opposite in Amazon's three datasets. The possible reason is that the data volume of Frappe and MovieLens is small, so the empirical error is dominant compared to the model complexity, which requires a larger $d$ to ensure a smaller empirical error. For the last three datasets with sufficient data, a small embedding size may already guarantee a small empirical error, and the model complexity turns out to be the major concern. Therefore, for the dataset with sufficient data, we suggest using a smaller embedding size to enjoy both generalization and computational efficiency.
\subsection{Robustness Evaluation}
\begin{figure}[htbp]
	\centering
	\includegraphics[width=0.95\linewidth]{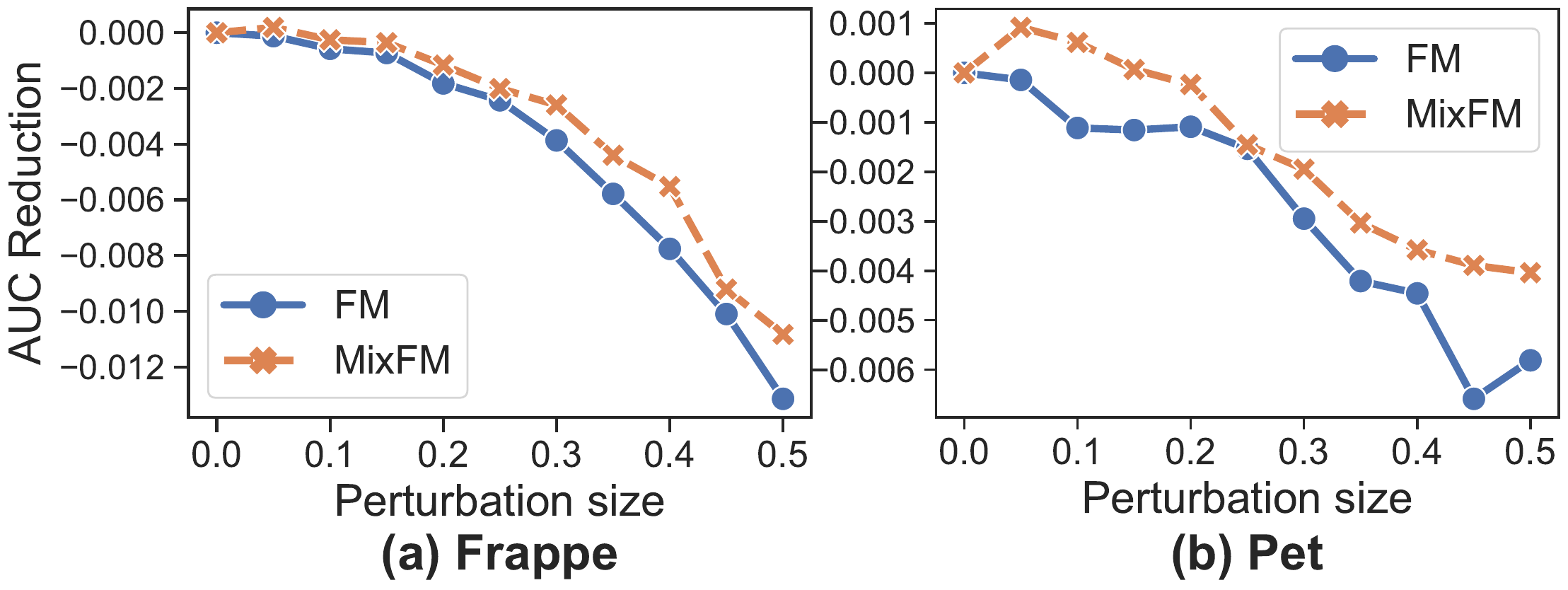}
	\caption{Performance under different perturbation sizes.}
	\label{fig: robust}
\end{figure}
Recent studies have shown that recommender systems are vulnerable \cite{he2018adversarial}; that is, adding small perturbations to the dataset can significantly reduce the recommendation quality. In this section, we test the performance of MixFM facing perturbations with different sizes. Fig. \ref{fig: robust} shows the model's performance degradation when random noises of different sizes are added to the Frappe and Electronics (we have similar results on other datasets). Firstly, we observe that the FM is fragile, and its performance will reduce significantly with the increase of noise intensity. Secondly, the AUC reduction on Electronics is smaller than that of Frappe. The possible reason is that Electronics has more data, corresponding to the finding that model robustness requires more data \cite{wu2021fight}. Finally, compared with FM, MixFM has a smaller reduction on AUC, revealing that using additional mixed data helps to improve FM's robustness.

\subsection{Training Runtime}
\begin{table}[htbp]
	\renewcommand\arraystretch{1.05}
	\centering
	\caption{Training Time (s)}
	\setlength{\tabcolsep}{0.015\linewidth}{
	\begin{tabular}{c|ccccc}
		\hline\noalign{\smallskip}
		& Frappe & MovieLens & Pet   & Electronics & Books \\
		\noalign{\smallskip}\hline\noalign{\smallskip}
		FM  & 88.1  & 220.5 & 189.3 & 497.2 & 657.9 \\
		CopyFM & 69.6  & 255.6 & 300.1 & 900.1 & 1230.3 \\
		MixFM & 168.7 & 569.9 & 333.1 & 902.0   & 1117.1 \\
		SMFM  & 187.5 & 592.0   & 340.8 & 951.2 & 1432.9 \\
		\noalign{\smallskip}\hline
	\end{tabular}%
}
	\label{tab:runtime}%
\end{table}%
Since the proposed strategies do not change models, no delay in inference occurs, which is critical in real-world applications. But it is undeniable that the training time complexity is linearly proportional to the dataset size. Table \ref{tab:runtime} shows the 50-round training time (in seconds) when MixFM and SMFM double the data, so does the training time, which is consistent with the analysis. Notably, Fig. \ref{fig: ratio} shows that a considerable improvement actually requires fewer mixed samples(e.g., Frappe only supplemented 10\% augmented data in NFM), while Fig. \ref{fig: training_logloss} shows that the proposed augmentation approaches are easier to converge. In summary, the actual training delay is much smaller, which is tolerable.

\section{Conclusion}
In this paper, we first developed a Mixup powered Factorization Machine (MixFM), a domain-knowledge-independent approach for enhancing FMs. Specifically, MixFM could efficiently generate additional data to assist FM in learning feature interactions. Second, we revealed the importance of generating informative samples for model decisions and further put forward a Saliency-guided Mixup powered Factorization Machine (SMFM), which generates more informative mixed samples via the customized saliency function. 
Third, we delivered the first generalization bound for FM, revealing the role of data and embeddings on model performance. At the same time, we proved that MixFM minimizes the generalization error upper bound, which theoretically guarantees the effectiveness of the proposed strategy. Finally, we conducted extensive experiments on multiple real-world datasets, and the results demonstrated the effectiveness of our proposals. 
Note that the proposed ideas are decoupled from the model, and theoretically, they can be applied to any differentiable model. Although we verified the potential on FM variants,
in future work, we will extend them to more recommendation models.


%

\section{Detailed Proofs}
In this section, we give detailed proofs of the generalization error upper bound for FM and MixFM, respectively.

\label{proofs}
\noindent\textbf{Proof of Theorem \ref{theorem: rademacher_fm}.}
The function family of FM is defined as Eq. \ref{problem: fm}, that is $\mathcal{H}_{\gamma}:=\{f:u\mapsto \theta^Tu+w_0, \left\|\theta\right\|^2\le\gamma^2\}$. $\mathcal{D}=\{u^i\}_{i=1}^n$ when we convert $x^i$ to $u^i$ according to Definition \ref{def: linear_fm}. Then, the empirical Rademacher complexity of the FM with respect to the dataset $\mathcal{D}$ is defined as:
\begin{equation}
\begin{split}
\hat{\Re}_{\mathcal{D}}(\mathcal{H}_{\gamma})&=\mathbb{E}_\sigma\left[\sup\limits_{\left\|\theta\right\|^2\le\gamma^2}\frac{1}{n}\sum_{i=1}^n\sigma_i(\theta^Tu^i+w_0)\right]\\
&=\frac{1}{n}\mathbb{E}_\sigma\left[\sup\limits_{\left\|\theta\right\|^2\le\gamma^2}\theta^T\sum_{i=1}^n\sigma_iu^i\right]\\
&\le \frac{\gamma}{n}\mathbb{E}_\sigma\left[\left\|\sum_{i=1}^n\sigma_iu^i\right\|\right]\\
&\le \frac{\gamma}{n}\left[\mathbb{E}_\sigma\left[\left\|\sum_{i=1}^n\sigma_iu^i\right\|^2\right]\right]^{1/2}\\
&=\frac{\gamma}{n}\left[\mathbb{E}_\sigma\left[\sum_{i,j=1}^n\sigma_i\sigma_j(u^i)^Tu^j\right]\right]^{1/2}\\
&\le \frac{\gamma}{n}\left[\sum_{i=1}^n\left\|u^i\right\|^2\right]^{1/2}.
\end{split}
\nonumber
\end{equation}
Since $u=[\bigvee_{i=1}^{m-1}\bigvee_{j=i+1}^{m}R(x_ix_j,d), R(0,d^2m^2-dm(m-1)/2)]^T\in\mathcal{R}^{d^2m^2}$, and $\left\|x\right\|_{0}\le \tau$, then only $d\tau(\tau-1)/2$ dimensions of $u$ may not be 0. Given that $\left\|x\right\|_{\infty}\le 1$, then we have $\left\|u\right\|^2\le \frac{d\tau(\tau-1)}{2}$. Therefore,
$$\hat{\Re}_{\mathcal{D}}(\mathcal{H}_{\gamma})\le \sqrt{\frac{\gamma^2d\tau(\tau-1)}{2n}}.$$
The proof is complete.

\noindent\textbf{Proof of Corollary \ref{theorem: generalization_fm}.}
We first introduce a theorem to support the proof.
\begin{theorem}[\cite{mohri2018foundations}]
	\label{theorem: generalization}
	Let $\mathcal{H}$ be a family of functions mapping from any sample to $[0,1]$. Then, for any $\delta>0$, with probability as least $1-\delta$ over the draw of an i.i.d. dataset $\mathcal{D}$ of size $n$, the following holds for all $f\in\mathcal{H}$:
	$$R_{\mathcal{D}}(f)\le\hat{R}_{\mathcal{D}}(f)+2\hat{\Re}_{\mathcal{D}}(\mathcal{H})+3\sqrt{\frac{\log \frac{2}{\delta}}{2n}},$$
	where $R_{\mathcal{D}}(f)$ and $\hat{R}_{\mathcal{D}}(f)$ represent the generalization error and empirical error of the model $f$, respectively.
\end{theorem}
In FM, the hypothesis space is $\mathcal{H}_{\gamma}$, and Theorem \ref{theorem: rademacher_fm} gives the empirical Rademacher complexity of $\mathcal{H}_{\gamma}$. By substituting them into Theorem \ref{theorem: generalization}, Corollary \ref{theorem: generalization_fm} can be proved.

\noindent\textbf{Proof of Theorem \ref{theorem: regularization}.}
We first provide a related finding to simplify our proof.
\begin{lemma}[\cite{zhang2020does}]
	\label{theorem: mixup_simply}
	Suppose $\mathcal{D}=\{(x^i,y^i)\}_{i=1}^n$ is the dataset of size $n$, Let $\widetilde{\mathcal{D}}=\{(\tilde{x}^i,\tilde{y}^i)\}_{i=1}^n$, where $\tilde{x}^i=\lambda x^i+(1-\lambda)r_x$, $\tilde{y}^i=\lambda y^i+(1-\lambda)r_y$, $\lambda\sim Beta(\alpha,\beta)$, and $(r_x,r_y)\sim \mathcal{D}$. If the loss function has the form similar to $l(x,y)=h(f(x))-yf(x)$, where $f(x)$ is the predict of $x$, and $h(\cdot)$ is an arbitrary function, then 
	\begin{equation}
	    \begin{split}
	        &\mathbb{E}_{\lambda\sim Beta(\alpha,\beta)}l(\tilde{x}^i,\tilde{y}^i)\\
	        =&\mathbb{E}_{\lambda\sim \frac{\alpha}{\alpha+\beta}Beta(\alpha+1,\beta)+\frac{\beta}{\alpha+\beta}Beta(\beta+1,\alpha)}l(\tilde{x}^i,y^i).
	    \end{split}
	\end{equation}
	
\end{lemma}

Given the difficulty of analyzing the Mixup in $x$ and $y$ simultaneously, Lemma \ref{theorem: mixup_simply} points out that if the loss function is $l(x,y)=h(f(x))-yf(x)$, then performing Mixup on the data pair $(x,y)$ can be simplified just to perform Mixup on the $x$, but $\lambda$ follows $\frac{\alpha}{\alpha+\beta}Beta(\alpha+1,\beta)+\frac{\beta}{\alpha+\beta}Beta(\beta+1,\alpha)$ instead of $Beta(\alpha,\beta)$.

Now we apply this lemma to MixFM. In the paper, we study the logistic loss $$l(x,y)=\log (1+\exp(f(x)))-yf(x).$$ 

Let $h(x)=\log (1+\exp(f(x)))$, then we can directly apply Lemma \ref{theorem: mixup_simply}, and only need to consider the Mixup on $x$ by setting $\lambda\sim \frac{\alpha}{\alpha+\beta}Beta(\alpha+1,\beta)+\frac{\beta}{\alpha+\beta}Beta(\beta+1,\alpha)$. For simplicity, we define it as $\tilde{\lambda}$. In the following proof, we will focus on the linear form of FM in Difinition \ref{def: linear_fm}; that is, $f(u)=\theta^Tu+w_0$.

Since $u=[\bigvee_{i=1}^{m-1}\bigvee_{j=i+1}^{m}R(x_ix_j,d), R(0,d^2m^2-dm(m-1)/2)]^T$, according to Eq. \ref{eq: mix}, we have the mixed data as follows: $$\tilde{u}=\left[\bigvee_{i=1}^{m-1}\bigvee_{j=i+1}^{m}R(\tilde{x}_i\tilde{x}_j,d), R(0,d^2m^2-dm(m-1)/2)\right]^T.$$ 

Considering each term $\tilde{x}_i\tilde{x}_j$, we have $\mathbb{E}_{r_x}\tilde{x}_i\tilde{x}_j=\tilde{\lambda}^2x_ix_j+\tilde{\lambda}(1-\tilde{\lambda})\left(x_i(r_x)_j+(r_x)_ix_j\right)+(1-\tilde{\lambda})^2(r_x)_i(r_x)_j$. Notice that we study centralized data $\mathbb{E}_{r_x\sim\mathcal{D}}(r_x)_i=0$, so $$\tilde{u}=\tilde{\lambda}^2 u+(1-\tilde{\lambda})^2r_u,$$ where $$r_u=\left[\bigvee_{i=1}^{m-1}\bigvee_{j=i+1}^{m}R((r_x)_i(r_x)_j,d), R(0,d^2m^2-\frac{dm^2-dm}{2})\right]^T.$$
Therefore, performing traditional Mixup on $x$ can be regarded as performing a mixture on $u$ to get $\tilde{u}$, and the mixing rule is $\tilde{u}=\tilde{\lambda}^2 u+(1-\tilde{\lambda})^2r_u$. 
For the model $f(\tilde{u})=\theta^T\tilde{u}+w_0$, the prediction is invariant to the shifting and scaling of $\tilde{u}$, so we study training the model on the data $\tilde{u}=\frac{1}{\tilde{\lambda}^2}[\tilde{\lambda}^2 u+(1-\tilde{\lambda})^2r_u]$.

The loss after Mixup is $l(\tilde{u},y)=\mathbb{E}_{\tilde{\lambda},r_u}[h(\tilde{u}^T\theta+w_0)-y(\tilde{u}^T\theta+w_0)]$, where $h(x)=\log (1+\exp(x))$. Using the second-order approximation of $l(\tilde{u},y)$ and setting $f(u)=u^T\theta+w_0$, we have
\begin{equation}
\begin{split}
&\mathbb{E}_{\tilde{\lambda},r_u}[h(f(\tilde{u}))-yf(\tilde{u})]
\approx\mathbb{E}_{\tilde{\lambda},r_u}[h(f(u))-yf(u)\\
&+(h'(f(u))-y)(\tilde{u}-u)^T\theta
+h''(f(u))\theta^T(\tilde{u}-u)(\tilde{u}-u)^T\theta].
\end{split}
\nonumber
\end{equation}

Among them, $\mathbb{E}_{\tilde{\lambda},r_u}(\tilde{u}-u)=\textbf{0}$, $h'(x)=g(x)$, $h''(x)=g(x)(1-g(x))$, where $g(\cdot)$ is the Sigmoid function. $\mathbb{E}_{\tilde{\lambda},r_u}(\tilde{u}-u)(\tilde{u}-u)^T=\mathbb{E}_{\tilde{\lambda}}\frac{(1-\tilde{\lambda})^4}{\tilde{\lambda}^4}\frac{1}{n}\sum_{i=1}^nu^i(u^i)^T$. As a result,
\begin{equation}
\begin{split}
&\mathbb{E}_{\tilde{\lambda},r_u}[h(f(\tilde{u}))-yf(\tilde{u})]
\approx\mathbb{E}_{\tilde{\lambda},r_u}[f(u))-yf(u)\\
&+g(f(u))(1-g(f(u)))\frac{(1-\tilde{\lambda})^4}{\tilde{\lambda}^4}\theta^T\hat{\Sigma}_u\theta],
\end{split}
\nonumber
\end{equation}
where $\hat{\Sigma}_u=\frac{1}{n}\sum_{i=1}^nu^i(u^i)^T$.
Thus,
\begin{equation}
\nonumber
\begin{split}
&\mathcal{L}(\widetilde{\mathcal{D}})=\frac{1}{n}\sum_{i=1}^{n}l(\tilde{u}^i,y)
=\mathcal{L}(\mathcal{D})\\
&+\frac{1}{n}\sum_{i=1}^{n}g(f(u))(1-g(f(u)))\mathbb{E}_{\tilde{\lambda}}\frac{(1-\tilde{\lambda})^4}{\tilde{\lambda}^4}\theta^T\hat{\Sigma}_U\theta,
\end{split}
\end{equation}
where $\tilde{\lambda}\sim \frac{\alpha}{\alpha+\beta}Beta(\alpha+1,\beta)+\frac{\beta}{\alpha+\beta}Beta(\beta+1,\alpha)$.
To sum up, the theorem is proved.

\noindent\textbf{Proof of Theorem \ref{theorem: generalization_mixfm}.}
The function family of FM training by Mixed data is defined as $$\widetilde{\mathcal{H}}_{\gamma}:=\{f:u\mapsto \theta^Tu+w_0, \mathbb{E}_ug(f(u))(1-g(f(u)))\theta^T{\Sigma}_u\theta\le\gamma\}.$$ 

Since $f\in\widetilde{\mathcal{H}}_{\gamma}$ maps any sample to $[0,1]$, we have $g(f(u))(1-g(f(u)))\ge\frac{e}{(1+e)^2}$, and then $\theta^T{\Sigma}_u\theta\le\frac{(1+e)^2\gamma}{e}$. Let $\hat{u}^i=\Sigma_u^{\dagger/2}u^i$, $\hat{\theta}=\Sigma_u^{1/2}\theta$, then $\hat{\left\|\theta\right\|}^2=\theta^T{\Sigma}_u\theta\le\frac{(1+e)^2\gamma}{e}$. According to the definition of empirical Rademacher complexity, we have
\allowdisplaybreaks[4]
\begin{eqnarray}
\begin{split}
\hat{\Re}_{\mathcal{D}}(\widetilde{\mathcal{H}}_{\gamma})&=\mathbb{E}_\sigma\left[\sup\limits_{g(f(u))(1-g(f(u)))\theta^T{\Sigma}_u\theta\le\gamma}\frac{1}{n}\sum_{i=1}^n\sigma_i(f(u^i))\right]\\
&=\mathbb{E}_\sigma\left[\sup\limits_{\hat{\left\|\theta\right\|}^2\le\frac{(1+e)^2\gamma}{e}}\frac{1}{n}\sum_{i=1}^n\sigma_i\hat{\theta}^T\hat{u}^i\right]\\
&=\frac{1}{n}\mathbb{E}_\sigma\left[\sup\limits_{\hat{\left\|\theta\right\|}^2\le\frac{(1+e)^2\gamma}{e}}\hat{\theta}^T\sum_{i=1}^n\sigma_i\hat{u}^i\right]\\
&\le \sqrt{\frac{(1+e)^2\gamma}{en^2}}\mathbb{E}_\sigma\left[\left\|\sum_{i=1}^n\sigma_i\hat{u}^i\right\|\right]\\
&\le \sqrt{\frac{(1+e)^2\gamma}{en^2}}\left[\mathbb{E}_\sigma\left[\left\|\sum_{i=1}^n\sigma_i\hat{u}^i\right\|^2\right]\right]^{1/2}\\
&\le \sqrt{\frac{(1+e)^2\gamma}{en^2}}\left[\sum_{i=1}^n(\hat{u}^i)^T\hat{u}^i\right]^{1/2}\\
&=\sqrt{\frac{(1+e)^2\gamma}{en^2}}\cdot\sqrt{n\cdot rank(\Sigma_u)}.
\end{split}
\nonumber
\end{eqnarray}
As shown in the proof process of Theorem \ref{theorem: rademacher_fm}, only $d\tau(\tau-1)/2$ dimensions may not be 0. Besides, noting the definition of $u$, each value is repeated $d$ times, so $rank(\Sigma_u)\le\frac{\tau(\tau-1)}{2}$. Therefore, 
$$\hat{\Re}_{\mathcal{D}}(\widetilde{\mathcal{H}}_{\gamma})\le \sqrt{\frac{(1+e)^2\gamma\cdot \tau(\tau-1)}{2en}}.$$

Similar to the proof of Corollary \ref{theorem: generalization_fm}, we substitute $\hat{\Re}_{\mathcal{D}}(\widetilde{\mathcal{H}}_{\gamma})$ into Theorem \ref{theorem: generalization}, and Theorem \ref{theorem: generalization_mixfm} will be proved.

\section*{Acknowledgment}

The work was supported by grants from the National Natural Science Foundation of China (No. 62022077 and 61976198).

\ifCLASSOPTIONcaptionsoff
  \newpage
\fi


\bibliographystyle{IEEEtran}
\bibliography{ref}

\end{document}